\documentclass[aps,twocolumn,amsmath,amssymb,%
                pre,showpacs]{revtex4}

\usepackage[dvips]{graphicx}
\usepackage{epsfig}

\begin{document}

\title{Infinite Invariant Density Determines Statistics of Time Averages 
for Weak Chaos 
 }

\author{N. Korabel}
\author{E. Barkai}
\affiliation{Department of Physics, Institute of Nanotechnology and Advanced Materials, Bar Ilan University, Ramat-Gan
52900, Israel}

% 05.90.+m  Other topics is statistical physics, thermodynamics,
%            and nonlinear dynamical systems.
% 05.45.Ac Low dimensional chaos
% 74.40.De Chaos and Noise
\begin{abstract} 
Weakly chaotic non-linear maps with marginal fixed points
have an infinite invariant measure.
 Time averages of integrable and non-integrable
observables remain random even in the long time limit. 
Temporal averages of integrable observables are described by the 
Aaronson-Darling-Kac theorem. We find the distribution 
of time averages of  non-integrable observables, for example
the time average position of the particle, $\overline{x}$. We show
how this distribution is related to the infinite invariant density. 
We establish
four identities between amplitude
ratios controlling the statistics of the problem. 
\end{abstract}
\maketitle

 Low dimensional
chaotic systems by definition have positive Lyapunov exponents
and have been extensively used to test basic assumptions of statistical
physics. Weakly chaotic systems have zero Lyapunov exponents,
namely the separation of trajectories is sub-exponential, 
though the deterministic
motion remains quasi-random. In many cases discrete maps 
are used to model the dynamics, since they help
to establish a deep understanding of the fundamental issues without being too
complicated (importantly numerics 
converge faster than in more realistic models).
In particular Pomeau-Manneville \cite{PM} maps
are weakly chaotic \cite{Gaspard} and are characterized by marginal 
instability. 
These maps were used
to model intermittency \cite{PM}, anomalous diffusion \cite{Gei84,Gei85,ZK93,Artuso} 
and aging \cite{BarkaiAging}.
Such systems
are described by an infinite invariant density ($\infty$D) \cite{Aaronson,Tasaki}: a non-normalizable
density defined below. 
It is also well known that temporal averages 
in such systems  are
not equal to a corresponding ensemble average, instead time averages
 remain random
variables even in the long measurement time limit
\cite{Aaronson,Thaler02,TZ06,Golan,Akimoto08}. 
Since chaos is 
a precondition for statistical physics, it is not very surprising that
weak chaos implies the breakdown of standard ergodic theory.  

 For an ergodic process, in the long time limit the time average of an
 observable
is equal to the corresponding ensemble average.
The ensemble average and hence the time average are obtained from the
 normalized invariant density, if it exists. 
A fundamental extension of standard
 ergodic theory is to find the distribution
of time averages of generic observables for weakly chaotic systems
where the underlying invariant density is non-normalizable. 
 The Aaronson-Darling-Kac (ADK)
theorem \cite{Aaronson} gives a partial answer to this problem. Briefly, 
an observable whose average with respect
to  the $\infty$D is finite, the distribution  of properly
scaled temporal averages is the
Mittag-Leffler distribution. The $\infty$D 
is essential for the description of these fluctuations.
For example, the separation of trajectories is described by a
 stretched exponential
(a manifestation of weak chaos) and the distribution of separation
rates is provided by the ADK theorem \cite{Kor09,Akimoto10}. 

In this manuscript, we 
consider the very large class of non-integrable observables. 
We focus on the position of
a particle $x_t$ in an interval $(0,L)$ and obtain the distribution of
its time average. 
Importantly we show how the distribution
of time averages of
non integrable observables is related to the underlying $\infty D$. 
 Previously Thaler and Zweim\"uller \cite{Thaler02,TZ06} 
considered an important 
non-integrable observable: the occupation fraction; i.e., the fraction of
time the particle spends within a given domain. They rigorously
 showed it is 
described by the Lamperti distribution (see details below). We provide a
very general conjecture for the distribution of time averages
of non-integrable observables,
 without giving a rigorous proof
but rather relying on simple arguments. Further we derive the
identity of  four
amplitude ratios which govern the
statistics of the problem. These identities bridge
between  the stochastic and dynamical theories in this field. 

%----------------------------------------------------------------------
 % Fig. 1
 %----------------------------------------------------------------------
 \begin{figure}
 %\vspace{-20pt}
 \centerline{\psfig{figure=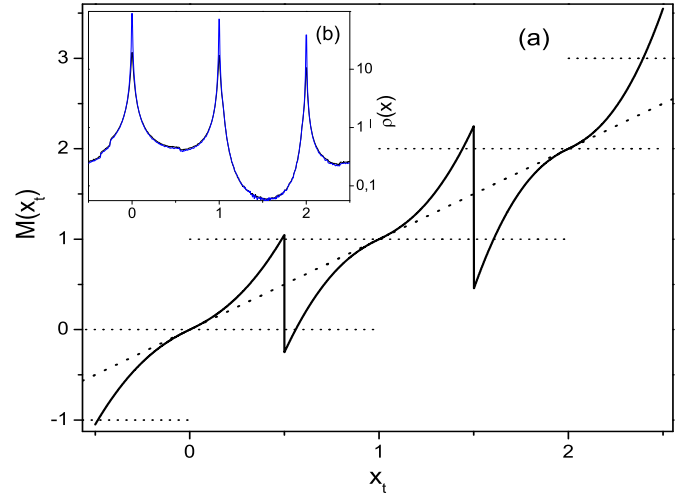,width=100mm,height=80mm}}
 \caption{(Color online) (a) The map Eq. (\ref{eq18}) 
with $L=3$ and $\alpha=3/4$
has IFPs on $0,1,2$. (b) The $\infty$D exhibits non-integrable
divergence on the IFPs (parameters are
given in third example in the text). }
%parameters $\alpha=0.75$, $\bar{a}_0=1.1$, $\bar{a}_1=1.5$, $\bar{a}_2=2.1$. 
%Arrows indicate direction of trajectories from pre-image points $c_i$ to corresponding MSFP $P_i$ ($i=0, 1, 2$). (b): Scaled density of the map $M_L(x_t)$, $t^{1-\alpha} \rho(x_t,t)$, calculated at $t=10^{3}$ and $t=10^{4}$ using $10^7$ trajectories uniformly distributed at $t=0$. In the long time limit the scaled density $t^{1-\alpha} \rho(x_t,t)$ converges to the infinite density $\bar{\rho}(x_t)$ Eq.\ (\ref{inf_rho}). (c): Density of injection points $P^{in}(x_t)$ calculated at $t=10^{2}$ and $t=10^{3}$ using $10^7$ trajectories.}
 \label{Fig1}
 \end{figure}
 %----------------------------------------------------------------------

{\em Model and observable.}
 We consider measure preserving maps $x_{t+1}  = M(x_t)$ with $x_t \in (0,L)$. 
Our observable is $x_t$ and our goal is to calculate the distribution
of its time average  $\overline{x} = \sum_{t=0}^{t-1} x_t / t$,
in the limit of long time. We assume that the map has $N$ indifferent fixed
points (IFPs) located on
$\{ x(1), ... x(j) , ... x(N) \}$ such that 
$M(x) \sim x + 2^{1/\alpha} a_j | x- x(j)|^{1 + 1/\alpha}$ as $x\to x(j)$
and $a_j \neq 0$ (IFPs are also called marginal fixed points).
Throughout this work, $j$ is a label of the IFPs. Here we consider $0<\alpha<1$
since in that regime the distribution of $\overline{x}$ is non-trivial. 
An example map is shown in Fig. \ref{Fig1}. We consider
maps where the trajectory of the particle visits the vicinity of
all the IFPs; i.e., we exclude stable points or a decomposable
phase space, so the transformation has an
infinite invariant measure. Such maps exhibit non-Gaussian
intermittency and hence have attracted vast research
using various methods such as CTRW \cite{Gei84,Gei85,ZK93}
and periodic orbit theory \cite{Artuso}.

 {\em Power law sojourn times are related to the injection probability}.
Let us consider the IFP $x(1)$ which we designate to be on the origin
$x(1) = 0$. In the vicinity of this point the map is
$x_{t+1} \simeq x_t +  2^{1/\alpha} a_1 (x_t)^{1/\alpha + 1} $
for $x_t>0$ and $0<\alpha<1$ while $a_1>0$ (other IFPs have constants
$a_j$).
Starting on $x_0$ the time $\tau$  it takes the particle to reach a threshold
$x_c$ is determined by the continuous approximation of the map
$dx/dt\simeq 2^{1/\alpha} a_1 (x)^{1/\alpha + 1} $. Following
Geisel and co-workers \cite{Gei84} this gives
\begin{equation}
\tau = \alpha { \left( x_0 \right)^{ - 1/\alpha} - \left(x_c\right)^{-1/\alpha} \over
2^{1/\alpha} a_1 } .
\label{eq01} 
\end{equation}
During the evolution the particle is injected in the vicinity of the
IFP many times and hence $x_0$ is  treated
as a random variable whose probability density function (PDF)
 is $P^{{\rm in}} [x_0]$. 
It follows that the waiting time $\tau$, the time the particle remains
in the vicinity of the  $j=1$ IFP, is a random variable with the PDF
$\psi_1 ( \tau ) = P^{{\rm in}} [x_0] |d x_0 / d \tau|$.
A similar formula holds for  the $j$th IFP. 
 Using Eq. (\ref{eq01}) one finds the PDF of
waiting times \cite{Gei84}
\begin{equation}
\psi_j (\tau) \sim  A_j \tau^{- (1 + \alpha) }\  
\mbox{with} \
A_j = P^{{\rm in}} \left[ x(j) \right]  {\alpha^{1 + \alpha}  \over 2 | a_j|^{\alpha}}. 
\label{eq02}
\end{equation}
Notice that this result is independent of the threshold $x_c$.
Here it is assumed that the injection PDF $P^{{\rm in}}[x(j)]$
 is smooth in the vicinity of the 
IFP.
Eq. (\ref{eq02}) is well known but actually rather formal since it expresses
$\psi_j(\tau)$ in terms of the unknown injection PDF. Below we
will relate the injection PDF with the $\infty$D. 
The power-law PDF Eq. (\ref{eq02}) indicates a diverging
average sojourn time since $0<\alpha<1$. 
 The corresponding 
stochastic picture \cite{Gei84,ZK93}  is of a particle jumping between
neighborhoods of the IFPs $\{ x(1) \cdots x(N) \}$ with
power law sojourn times for the trapping events.

{\em The infinite invariant density} plays a crucial role and it is defined
as \cite{remarkR}
\begin{equation}
\overline{\rho}(x) \simeq  \rho(x,t) / t^{\alpha -1} , \ \ \ \ \ t \to \infty 
\label{eq03}
\end{equation}
where $\rho(x,t)$ is the density of particles 
[in simulations we
use initial conditions uniformly spread in $(0,L)$]. 
When $\alpha<1$, the invariant density is non-normalizable,
$\int_0 ^L \overline{\rho}(x) {\rm d} x = \infty$, and hence its name.
Such $\infty$Ds are not common in physics though
recently an application was suggested in the context of cooling in
optical lattices \cite{Kessler}.  
Note that the density $\rho(x,t)$ is, as usual,
normalizable for any finite $t$ since the maps conserve the number 
of particles. In the vicinity of the IFP $x(j)$ one finds the 
non-integrable behavior 
\begin{equation}
\overline{\rho}(x) \simeq b_j | x- x(j) |^{-1/\alpha}, 
\label{eq04}
\end{equation}
where $b_j\geq 0$ is an amplitude which is generally unknown.
An example $\infty$D is shown in Fig. \ref{Fig1} 
based on a numerical simulation which allows us to estimate the $b_j$s.  

To understand better such a behavior we use simple arguments.
Note that the density normalized to unity is
\begin{equation}
\rho(x,t) {\rm d} x \simeq t_{x,x+ {\rm d} x} / t 
\label{eq05}
\end{equation}
where $t_{x,x+ {\rm d} x }$ is the time the particle spends in $(x,x+dx)$
\cite{remark1}. Let us consider the vicinity of the first
IFP $x(1) = 0$. The time $t_{x,x+{\rm d} x}$ 
is proportional to $N_R$: the number of injections
to the vicinity of the IFPs,  multiplied by $P^{{\rm in}}[x(1) ] {\rm d} x$
[which gives the number of visits in the interval $(x,x+dx)$].
$t_{x,x+dx}$ is also proportional to the time the particle
stays in $(x,x+dx)$ during each visit,
 which we call $\Delta t$. Thus, close
to the IFP, 
\begin{equation}
\rho(x,t) \simeq {  N_R P^{{\rm in}}\left[x(1)\right] \Delta t \over t} .
\label{eq06}
\end{equation}
As is well known from renewal theory \cite{Gei84,ZK93}
the number of renewals or injections scales like $N_R \simeq C t^\alpha$.
The pre-factor
$C$ can be roughly estimated however below we show that it is an irrelevant 
parameter. Using Eq.  (\ref{eq01}) we have when $x\to 0$
\begin{equation}
\Delta t \simeq { \alpha x^{-1/\alpha} \over 2^{1/\alpha} a_1 },
\label{eq07}
\end{equation}
so that the closer the particle is to the IFP $x(1)=0$ the longer is $\Delta t$.
Similarly we analyze other IFPs. Putting these pieces of information
together, we find
\begin{equation}
\rho(x,t) \simeq b_j 
{ |x- x(j)|^{-1/\alpha} \over t^{1- \alpha} } \ 
\mbox{where} \ 
b_j=  {C  \alpha P^{{\rm in}} \left[ x(j) \right]\over   2^{1/\alpha} a_j}.
\label{eq08}
\end{equation}
%\rho(x,t) \sim \underbrace{ c { \alpha P^{{\rm in}} \left[ x(j) \right] \over 2^{1/\alpha} a_j }}_{b_j}   { |x- x(j)|^{-1/\alpha} \over t^{1- \alpha} } \ \ \
%
%where $b_j=  C  \alpha P^{{\rm in}} \left[ x(j) \right]  2^{-1/\alpha}/a_j  $.
 Eq. (\ref{eq08}) shows, a  relation
between the amplitudes of the $\infty$Ds; i.e., the $b_j$s
and the injection probabilities $P^{{\rm in}}[x(j)]$ \cite{remark2}.

%----------------------------------------------------------------------
% Fig. 2
%----------------------------------------------------------------------
\begin{figure}
%\vspace{20pt}
\centerline{\psfig{figure=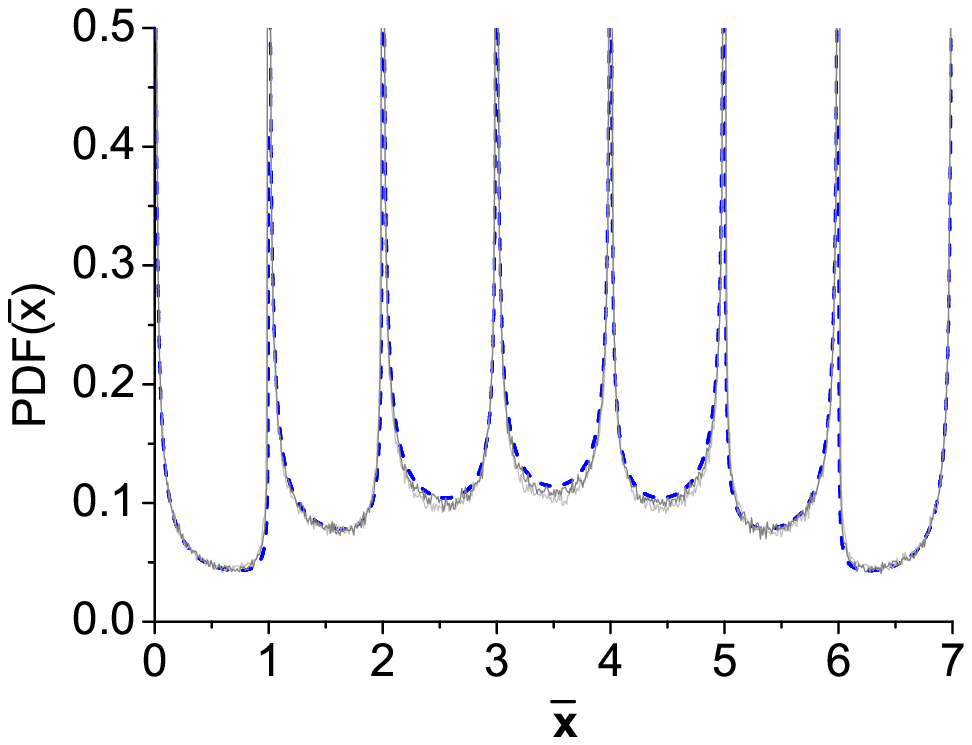,width=48mm,height=47mm} \hspace{-0.7cm}
\psfig{figure=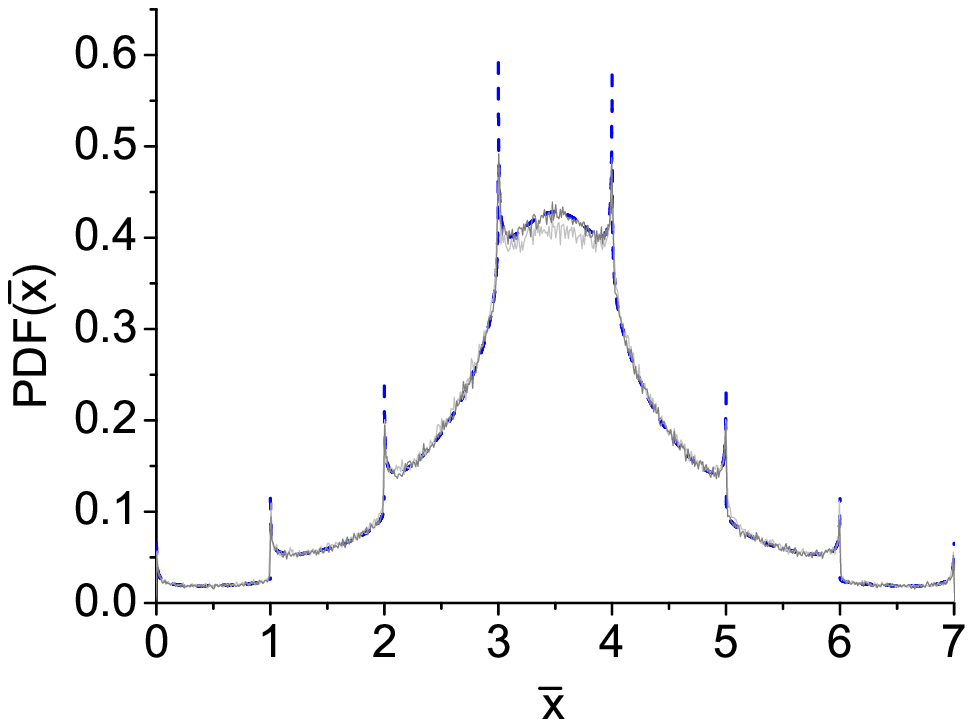,width=48mm,height=47mm}}
\caption{(Color online) 
Numerical PDF of $\overline{x}$ (solid line), 
perfectly
matches the theory Eq. (\ref{eq19}) (dashed line) without any fitting [map Eq. (\ref{eq18}), 
left panel
$\alpha=0.3$, right $\alpha=0.75$ and $t=10^6$].}
\label{Fig2}

\end{figure}
%----------------------------------------------------------------------

{\em The time average $\overline{x}$ } is now considered. During
the evolution the trajectory of the particle $x_t$ spends long times,
of the order of the measurement time in the vicinity of
the IFPs. In contrast the time it takes the particle to jump between one
IFP state to another is short and can be neglected. 
Hence along a trajectory
$x_t$ attains observable values which are (nearly) equal to the locations
of the IFP $\{ x(1) \cdots x(N)\}$. In each one of these states
the particle remains a time $t_j$ with $j=1,\cdots , N$
which is the occupation time of state $j$. It follows that the time
average is 
\begin{equation}
\overline{x} \simeq  {\sum_{j=1} ^N x(j) t_j \over t}  .
\label{eq09}
\end{equation} 
Each $t_j$ is a sum of many independent identically distributed random sojourn
times
drawn from the long tailed PDF $\psi_j(\tau)$, Eq. (\ref{eq02}). 
Hence  the occupation time $t_j$ is distributed
according to L\'evy statistics, i.e. the generalized central limit theorem
holds. More precisely $t_j$ is a stable random variable whose PDF is the
one sided L\'evy function with index $0<\alpha<1$.  Let $p_j^{{\rm eq}} = \langle t_j \rangle/t$ be the averaged occupation fraction treated rigorously
in
\cite{Thaler02,TZ06} 
 which is nothing but the probability that a member
of an ensemble of noninteracting particles is in the vicinity of 
the IFP $j$. Since the occupation time $t_j$
scale with $A_j$  and $t = \sum_{j=1} ^N t_j$ we get
\begin{equation}
p_j ^{{\rm eq}} = { A_j \over \sum_{j=1} ^N  A_j },
\label{eq10}
\end{equation} 
where $A_j$ is the amplitude of the waiting time PDF, defined in
Eq. (\ref{eq02}). 
Importantly, using  
Eqs. (\ref{eq02},\ref{eq10}),
\begin{equation}
p_j ^{{\rm eq}} = { P^{{\rm in}} \left[ x(j) \right] |a_j|^{-\alpha} \over
\sum _{j=1}^N P^{{\rm in}} \left[ x(j) \right] |a_j|^{-\alpha} },
\label{eq11} 
\end{equation} 
which relates occupation fractions with injection probabilities. 
Using Eqs. 
(\ref{eq08},\ref{eq11}) 
\begin{equation}
p_j ^{{\rm eq} } = { b_j |a_j|^{- \alpha + 1} \over \sum_{j=1} ^N b_j |a_j|^{-\alpha + 1 } },
\label{eq12} 
\end{equation} 
which relates the occupation fractions and 
the $\infty D$. 
 
 The distribution of observables like $\overline{x}$ was recently studied
within the continuous time random walk model, a stochastic approach
extensively applied, though so far without 
an underlying $\infty$D. Briefly, as mentioned,
$t_j$ is a stable random variable, and since $\overline{x}$ [Eq. 
(\ref{eq09})] is a linear combination of such independent random variables,
one finds the PDF of the time average  \cite{RB07}
\begin{equation}
f_\alpha\left( \overline{x} \right) = - { 1 \over \pi} \lim_{\epsilon \to 0} {\rm Im} { \sum_{j=1}^N p_j ^{{\rm eq}} | \overline{x} - x(j) + i \epsilon |^{\alpha -1} \over \sum_{j=1} ^N p_j ^{{\rm eq}} | \overline{x} - x(j) + i \epsilon|^\alpha},
\label{eq13} 
\end{equation}
where $i = \sqrt{-1}$ and ${\rm Im}$ denotes the imaginary part. 
We see that the PDF of $\overline{x}$ is controlled
by the nonlinearity of the map in the vicinity of the IFPs, i.e. $\alpha$,
the values of the observable on these points $\{ x(j) \}$, the equilibrium
probabilities $p^{{\rm eq}}_j$ which in turn depend on either the 
$\infty$D, Eq. (\ref{eq12}), or the injection PDF, Eq. (\ref{eq11}).
Thus once the invariant density is known
one may obtain full information on the fluctuations of the
time average of our observable. 
Notice that when $\alpha \to 1$, Eq. (\ref{eq13}) yields 
$\lim_{\alpha \to 1} f_\alpha \left(\overline{x} \right) \sim \delta(\overline{x} - \langle x \rangle)$ 
where $\langle x \rangle = \sum p_j ^{{\rm eq}} x(j)$ is the ensemble average. 
For a general non-integrable observable ${\cal O} (x_t) $, the
distribution of the time average 
$\overline{{\cal O}} = \sum_{t=0} ^{t-1} {\cal O} (x_t)/t$
 is $f_\alpha\left( \overline{{\cal O}}\right)$ as in Eq.  
(\ref{eq13}) where on the right hand side 
we replace $x(j)$ with  ${\cal O}(x(j))$. 

{\em A first illustration} will be a system with two IFPs. We consider
$x_t \in (0,1)$ and
\begin{equation}
M(x_t) = \left\{ 
\begin{array}{l l}
x_t + 2^{1/\alpha} (x_t)^{1 + 1/\alpha} & 0 < x_t <1/5 \\
1 + { 1/5 - x_t \over 7/20 } & 1/5 < x_t < 11/20 \\
x_t - 2^{1/\alpha} (1 - x_t)^{1 + 1/\alpha} & 11/20<x_t <1 ,
\end{array}
\right.
\label{eq14}
\end{equation}
hence $x(1) = 0$ and $x(2) = 1$ are the IFPs of the map and $|a_1|=|a_2|$.
We first concentrate on the injection PDF $P^{{\rm in}}[x]$. We partition the
map into two parts with a boundary on $0<x_c<1$. Following a trajectory
we record events where the particle jumps over the boundary,
either from left to right or vice versa. Each time the particle is
injected into one of the domains $x<x_c$ or $x>x_c$ we record its 
landing position and thus generate a histogram which gives $P^{{\rm in}}[x]$.
Not surprisingly $P^{{\rm in}}[x]$ will depend on the choice of $x_c$.
 However, interestingly, the ratio $P^{{\rm in}} [x(1) ]/P^{{\rm in}} [x(2)]$
is a constant independent of the value of $x_c$.
 To understand this behavior note
that according to 
Eq. (\ref{eq08}) we get the amplitude ratio relation
\begin{equation}
  b_2 / b_1 =  P^{{\rm in}} \left[ x(2) \right] / P^{{\rm in}} \left[ x(1) \right]
\label{eq15}
\end{equation} 
and since $b_2/b_1$ is clearly $x_c$ independent so is the right hand
side of this equation. Starting with a uniform density we evolve the system
until time $10^4$,  obtain an estimate for the $\infty$D
$\overline{\rho}(x)$, and with it find $b_1$ and $b_2$.  
For $\alpha =0.75$  we find $b_1=0.075, b_2 = 0.16$ and
for $x_c=0.5$ $P^{{\rm in}}[x(1)] = 0.86$ and  $P^{{\rm in }} [x(2) ] = 1.86$
while $P^{{\rm in}}[x(1)] = 1.18$ and  $P^{{\rm in }} [x(2) ] = 2.58$
for $x_c = 0.3$. Hence Eq. (\ref{eq15}) stands the numerical test. We have
also verified this equation with other parameters.

 After we get the amplitudes of the  infinite invariant density,
 $b_1$ and $b_2$, we may calculate $p^{{\rm eq}} _1$ and
$p^{{\rm eq}} _2$ and so using Eq. (\ref{eq13})  we find
the PDF of $\overline{x}$ 
\begin{equation} 
f_\alpha \left( \overline{x} \right) = { \pi^{-1} \sin \left( \pi \alpha \right){\cal R}  \overline{x}^{\alpha -1 } \left( 1 - \overline{x} \right)^{\alpha -1} \over 
{\cal R}^2 \left( 1 - \overline{x} \right)^{2 \alpha} + \left( \overline{x} \right)^{2 \alpha} + 2 {\cal R} \cos \pi \alpha \left( 1 - \overline{x} \right)^\alpha \overline{x}^\alpha },
\label{eq16}
\end{equation} 
which is  the Lamperti PDF. The same distribution 
was previously obtained for the occupation fraction
\cite{Thaler02,TZ06,Golan}. As pointed out by Akimoto \cite{Akimoto08}
this is not surprising since both observables are identical  
{\em on the IFPs} [for the occupation
fraction the observable  is the step function which is $1$ on $x(2)=1$ 
and zero on $x(1)=0$]. 
The parameter ${\cal R}$ is 
\begin{equation}
{\cal R} = {A_2 \over A_1} ={ P^{{\rm in}}[x(2)] \over P^{{\rm in}} [x(1)] } =
{b_2 \over b_1} = {p^{{\rm eq}} _2 \over p^{{\rm eq}} _1} .
\label{eq17}
\end{equation}
Hence one has four  amplitude ratios related to the waiting times, 
the injection probabilities, the $\infty$D and the population
probabilities which determine
the PDF of $\overline{x}$.
 {\em Amplitude ratios} can be easily generalized for the case of 
$N$ IFPs and for the case where the $a_j$s are not all equal
\begin{equation}
{p_j ^{{\rm eq}} \over p_i ^{{\rm eq}}} = { A_j \over A_i} = { |a_j|^{-\alpha} P^{{\rm in}} [x(j) ] \over |a_i|^{-\alpha}  P^{{\rm in}} [x(i)] } =
{b_j |a_j|^{-\alpha+1} \over b_i |a_i|^{-\alpha +1} }.
\label{eq17a}
\end{equation}

%----------------------------------------------------------------------
 % Fig. 3
%----------------------------------------------------------------------
 \begin{figure}[t]
\centerline{\psfig{figure=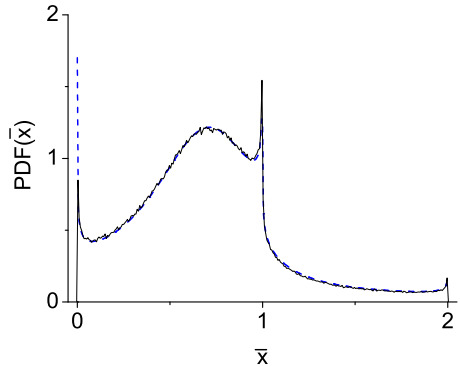,width=68mm,height=57mm} 
}
 \caption{(Color online) 
 Numerical simulations give the
PDF of $\overline{x}$ (solid line) that matches the analytical density (dashed line) 
Eq. (\ref{eq13})
 [map Eq. (\ref{eq18}), 
$t=10^6$, $L=3$ and $\alpha=3/4$].} 
 \label{Fig3}
 \end{figure}
%----------------------------------------------------------------------

{\em The second illustration} concerns maps with $N$ degenerate IFPs. 
We
consider $N=2 L$ with $L=8$ and $x_t \in (-1/2,7.5)$.
 The map is
\begin{equation}
M(x_t) = x_t +  \left\{
\begin{array}{l l}
2^{1/\alpha} \tilde{a}_k (x_t - k)^{1 + 1/\alpha} & k<x_t <k+1/2 \\
-2^{1/\alpha} \tilde{a}_k (-x_t + k)^{1 + 1/\alpha} & k-1/2<x_t <k,
\end{array}
\right.
\label{eq18}
\end{equation}
where $k=0,\cdots, L-1$. Here $16$ IFPs are on
$\{ x(1) = 0^{-},x(2)= 0^{+}, .... x(15)=7^{-}, x(16) = 7^{+}\}.$ 
We use periodic boundary conditions: if $x_t> 7.5$  or $x_t<-1/2$ we transform $x_t$ to $x_t - 8$ or $x_t + 8$ respectively. We set all $\tilde{a}_k=1$.
Then from symmetry we expect that all the amplitudes $b_j$ will be identical.
It then follows that $p^{{\rm eq}} _j = 1/(2 L)$. For this degenerate
case we get
\begin{equation}
f_\alpha \left( \overline{x} \right) = - {1 \over \pi} \lim_{\epsilon \to 0} 
{\rm Im} \sum_{j=0} ^{L-1} {\left( \overline{x} - j + i \epsilon\right)^{\alpha - 1}  \over \left( \overline{x} - j + i \epsilon\right)^\alpha } .
\label{eq19}
\end{equation}
Thus due to symmetry the distribution of $\overline{x}$ depends only on a
single parameter which is $\alpha$.   
 In Fig. \ref{Fig2} we show
the PDF of $\overline{x}$  
obtained numerically together with
theory Eq. (\ref{eq19}). For $\alpha=0.3$  the distribution
is wider than the case $\alpha=3/4$ since we expect
that as $\alpha \to 1$ the fluctuations will vanish. Notice that 
$f_\alpha(\overline{x})$ diverges on the
IFPs reflecting trajectories with a trapping time of the order
of the measurement time on one of these points.  

{\em The third example} is the map Eq. 
	(\ref{eq18}) with $L=3$ and hence IFPs are on 
$x(1)=0^{-},x(2)=0^{+},x(3)= 1^{-},x(4)=1^{+},x(5)= 2^{-},$ 
$x(6) = 2^{+}$. 
It follows that in the long time limit 
$\overline{x} \in (0,2)$. 
We use $\tilde{a}_1=1.1$, $\tilde{a}_2=1.5$ and $\tilde{a}_3 =2.1.$
We numerically obtain the $\infty$D for $\alpha = 3/4$
and estimate $b_1=b_2\simeq 0.059,
b_3=b_4\simeq 0.04, b_5=b_6\simeq 0.018$. Inserting these values
in Eq. (\ref{eq12}) 
we find:
$p^{{\rm eq}}_1=p^{{\rm eq}}_2 = 0.239$, 
$p^{{\rm eq}}_3 = p^{{\rm eq}}_4=0.175$ and 
$p^{{\rm eq}}_5=p^{{\rm eq}}_6 = 0.086$. 
This is compared with direct numerical computation
of the occupation fraction: 
$p^{{\rm eq}} _1 = p^{{\rm eq}} _2 \simeq 0.242, p^{{\rm eq}} _3 = p^{{\rm eq}} _4 \simeq 0.167, p^{{\rm eq}} _5= p^{{\rm eq}} _6\simeq 0.091 $. 
Deviations between the two methods are related to the divergence of the
$\infty D$ next to IFPs which induces errors in the
estimation of the $b_j$s. 
Inserting the latter values of $p^{{\rm eq}}_j$s  into Eq. 
(\ref{eq13}) we obtain the PDF of $\overline{x}$ which 
as shown in Fig. \ref{Fig3} perfectly matches direct numerical simulation.

 {\em Discussion.} 
We obtained
the distribution of time averages
of  non-integrable observables for
systems with IFPs with an infinite invariant measure. 
The $\infty D$, the occupation fractions, the injection probabilities,
and the amplitudes $A_j$ of the scale free distributions of the sojourn times,
 are all related  and can be used to determine the
non-trivial distribution of the temporal averages. 
There exists a vast number of physical systems with dynamics
governed
by power 
law trapping times 
similar to the maps under investigation.  
 A fundamental experimental 
question is whether such systems,
e.g. blinking quantum dots \cite{Barkai}, 
 two dimensional rotating flows \cite{Solomon,del}
and electro-hydrodynamic convection in liquid crystals \cite{Grigolini}
possess an infinite invariant measure. 
Hence it would be interesting to extract the invariant density
from the trajectories in such experiments.
If it is of infinite
measure, one could then use our theory 
to predict the distribution of
the temporal averages.  

{\bf Acknowledgement} 
This  work  was supported by the  Israel Science  Foundation.


\begin{thebibliography}{99}

\bibitem{PM}
Y.~Pomeau, and P.~Manneville, Commun. Math. Phys. {\bf 74}, 189 (1980); P.~Manneville, J.
Phys. {\bf 41}, 1235 (1980).

\bibitem{Gaspard}
P.~Gaspard,  and X.-J.~Wang,
Proc. Nat. Acad. Sci. USA, {\bf 85}, 4591 (1988).

\bibitem{Gei84} T.~Geisel, and S.~Thomae, Phys. Rev. Lett. {\bf 52}, 1936 (1984).

\bibitem{Gei85} T. Geisel, J. Nierwetberg and A. Zacherl, Phys. Rev. Lett. {\bf 54}, 616 (1985).

\bibitem{ZK93} G.~Zumofen, and J.~Klafter, Phys. Rev. E. {\bf 47}, 851 (1993).

\bibitem{Artuso} R. Artuso, P. Cvinatovi\'c, and G. Tanner,
Prog. Theor. Phys. Supl. {\bf 150}, 1 (2003). 

\bibitem{BarkaiAging}
E.~Barkai, Phys. Rev. Lett. {\bf 90}, 104101 (2003).

\bibitem{Aaronson} J.~Aaronson,
{\it An Introduction to Infinite Ergodic Theory},
(American Mathematical Society, 1997).

% Spectral properties of Piecewise Linear Intermittent Map 
\bibitem{Tasaki} S. Tasaki, and P. Gaspard,  J. of Statistical Physics 
{\bf 109} 
803 (2002).

\bibitem{Thaler02}
M.~Thaler,
%A limit theorem for sojourns near indifferent fixed points of one dimensional maps,
Ergod. Th. \& Dynam. Sys. {\bf 22}, 1289 (2002).

\bibitem{TZ06}
M.~Thaler, and R.~Zweimuller,
%Distributional limit theorems in infinite ergodic theory,
Probab. Theory Relat. Fields {\bf 135}, 15 (2006).

\bibitem{Golan} G. Bel, and E. Barkai, Europhysics Lett. {\bf 74} 12 (2006). 

\bibitem{Akimoto08}
%Generalized Arcsine Law and Stable Law in an Infinite Measure Dynamical System
T.~Akimoto, J. Stat. Phys. {\bf 132}, 171 (2008).

\bibitem{Kor09}
N.~Korabel, and E.~Barkai, Phys. Rev. Lett. {\bf 102}, 050601 (2009).
ibid, Phys. Rev. E. {\bf 82}, 016209 (2010).

\bibitem{Akimoto10}
T.~Akimoto,  and Y.~Aizawa, Chaos {\bf 20}, 033110 (2010).

\bibitem{remarkR} In the literature  the $\infty D$ is unique up to a
multiplicative
constant. Our definition Eq. 
(\ref{eq03}) is a convenient   choice. Multiplying the $\infty D$  with a 
constant does not alter our  main results, since they depend on ratios
of amplitudes. 

\bibitem{Kessler} D. A. Kessler,  and E. Barkai, Phys. Rev. Lett. {\bf 105},
 120602 (2010). 

\bibitem{remark1} Strictly this time is random for a given trajectory and we
mean an averaged time. 

\bibitem{remark2} Note that for any finite long time the non-integrable 
behavior is cut off \cite{Kor09}. 
Still, as we show in the text, the non-normalized
density plays a crucial role in the determination of the
distribution of the time averages through the $b_j$s. 

\bibitem{RB07}
A.~Rebenshtok and E.~Barkai,
%Distribution of Time-Averaged Observables for Weak Ergodicity Breaking,
Phys. Rev. Lett. {\bf 99}, 210601 (2007).
ibid
J. Stat. Phys. {\bf 133}, 565 (2008).

\bibitem{Barkai} F. D. Stefani, J. P. Hoogenboom, and E. Barkai,
Phys. Today {\bf 62}, 34 (2009). 

\bibitem{Solomon} T. H. Solomon, E. R. Weeks, and H. L. Swinney,
 Phys. Rev. Lett. {\bf 71}, 3975 (1993).

\bibitem{del} D. del-Castillo-Negrete,  Phys. of Fluids {\bf 10}, 576 (1998).

\bibitem{Grigolini} L. Silvestri, L. Fronzoni, P. Grigolini, and P. Allegrini,
Phys. Rev. Lett. {\bf 102}, 014502 (2009). 

\end{thebibliography}
\end{document}